# Hold On and Swipe: A Touch-Movement Based Continuous Authentication Schema based on Machine Learning

Jacob Mallet
Department of Computer Science
University of Wisconsin – Eau Claire
Eau Claire, WI, USA
e-mail: malletjc3227@uwec.edu

Laura Pryor
Department of Computer Science
University of Wisconsin – Eau Claire
Eau Claire, WI, USA
e-mail: pryorlk8701@uwec.edu

Dr. Rushit Dave
Department of Computer Science
University of Wisconsin – Eau Claire
Eau Claire, WI, USA
e-mail: daver@uwec.edu

Dr. Naeem Seliya
Department of Computer Science
University of Wisconsin – Eau Claire
Eau Claire, WI, USA
e-mail: seliyana@uwec.edu

Dr. Mounika Vanamala
Department of Computer Science
University of Wisconsin – Eau Claire
Eau Claire, WI, USA
e-mail: vanamalm@uwec.edu

Dr. Evelyn Sowells-Boone
Department of Computer Science
North Carolina A&T State University
Greensboro, NC, USA
e-mail: sowells@ncat.edu

*Abstract*— **In recent years, the amount of secure information being stored on mobile devices has grown exponentially. However, current security schemas for mobile devices such as physiological biometrics and passwords are not secure enough to protect this information. Behavioral biometrics have been heavily researched as a possible solution to this security deficiency for mobile devices. This study aims to contribute to this innovative research by evaluating the performance of a multi-modal behavioral biometric based user authentication scheme using touch dynamics and phone movement. This study uses a fusion of two popular publicly available datasets - the Hand Movement Orientation and Grasp (HMOG) dataset and the BioIdent dataset. This study evaluates our model's performance using three common machine learning algorithms; Random Forest, Support Vector Machine, and K-Nearest Neighbor reaching accuracy rates as high as 82%, with each algorithm performing respectively for all success metrics reported.**

## I. INTRODUCTION

Smartphones are increasingly becoming more integrated into our daily life. Online banking, E-commerce, and other services such as storing information on the cloud are becoming commonplace in society today. With this growing technology comes more and more personal information being stored on these devices, making it extremely important to keep this sensitive data behind secure authentication methods. One-time passwords, or physiological-based biometrics such as fingerprint and facial scanners are some of the most commonly used user authentication methods today. While these methods may be convenient for the user, they may also be susceptible to attackers waiting to steal any of the user's valuable data stored on the device. For example, one-time passwords can be stolen from the user, or attackers can imitate the user's fingerprint through smudge attacks. This looming threat on the user's sensitive information has been the motivation to explore improvements on the current security methods on devices. While there has been much research done on finding the best solution for mobile device security [1-3], utilizing behavioral biometrics as a layer of security is possibly one of the best solutions.

Behavioral biometrics make use of the user's behaviors during their interaction with the device. The precise yet unique differences in behavior while using a device is what is used to identify users from each other. Using behavioral biometrics as an authentication method can be secure as seen in [4-7]. Instead of having to remember a password to unlock a mobile device, a behavioral biometric system can authenticate a user by simply interacting with the device, as seen in [8,9], which can be much more convenient. While other biometric practices such as facial recognition or fingerprint scanner may capture data that is sensitive to the user, behavioral biometric data is much less intrusive. These factors make behavioral biometrics a prime candidate for future mobile authentication schemes.

In this paper, we propose a behavioral biometric authentication scheme using machine learning algorithms, specifically K-Nearest Neighbors (KNN), Support Vector Machine (SVM), and Random Forest (RF), to classify users using a dataset containing information about the user's interaction with the device. Our models utilize phone movement data along with touchscreen data to classify users, which has been seen to be an effective combination of biometrics as seen in [10-12]. The data was captured using the accelerometer, gyroscope, and magnetometer sensors, which are commonly found in many mobile devices. The novel contributions of this paper are as follows:

- Develop multiple machine learning algorithms, namely Random Forest (RF), K-Nearest Neighbors (KNN), and Support Vector Machine (SVM), to identify users based on their touch dynamics and touch movement data.
- Evaluate the performance of our classifiers when identifying users using a multi-modal model

## II. RELATED WORKS

Behavioral biometrics has been recently gaining much more attention in authentication schemes on smartphones. Many machine learning algorithms have been deployed to aid researchers in this domain as they are exceedingly capable in detecting the subtle patterns of a user's movement while identifying important features from a large raw dataset. A majority of the current research done on behavioral biometrics has fallen into one of two categories: touch dynamics [13, 14] or phone movement [15]. Touch dynamics commonly involves extracting features from a user's swipe, such as x and y coordinates, finger area, or pressure. Phone movement research leverages data collected using the accelerometer, gyroscope, and magnetometer sensors to identify features that can authenticate a user. After we conducted previous research [16, 17], we found that combining touch dynamics and phone movement to form a multi-modal model can increase accuracy.

In [18] researchers also decide to authenticate users using a multi-modal model, by combining touchscreen and sensor data. The model utilizes an application to capture the behavioral data from the user. The app involves giving user's a specific path to connect different points that are distributed across the screen randomly using swipe gestures. Additionally, this study also explores the minimum number of swipes needed to accurately distinguish users from each other. To do this, the model will evaluate data containing 1, 3, 5, 7 swipes. At first, [18] decided to only test their model just using their touchscreen data. They chose 59 features, which included gesture length, duration, acceleration, angular velocity and more. After running their data through classifiers Random Forest (RF), Support Vector Machine (SVM), K-Nearest Neighbor (KNN), Naïve Bayes, and XGBoost, they concluded that RF and XGBoost were the only models producing satisfactory results. They also concluded that three swipes was sufficient enough to identify a user. After doing this they decided to test RF, SVM, and XGBoost against attackers, while now giving their classifiers access to the data from the accelerometer. After preprocessing their touchscreen and accelerometer data, some of their top performing touch data features included mean, max, median, minimum, quartile one, quartile three, and the interquartile range of touch minor data. The swipe length, curvature, and duration all also performed well. Features that ranked at the top from the accelerometer sensor included the mean, quartile one, and minimum in the y directions, as well as the mean in the x direction and the quartile three in the z direction. After applying hyperparameter tuning and recursive feature elimination on just RF, [18] found that it made very little difference to the results, yet was very costly. Finally, their RF classifier performed the best, producing a 1.40% False Acceptance Rate and a 2.08% False Rejection Rate. This paper highlighted that taking advantage of touch dynamics as well as phone movement produces better results.

The authors of [19] similarly use a multi-modal system using the accelerometer and touchscreen data. They also explored the effect on their model if the user was reading or navigating on the phone, and if the user was walking or sitting while using the phone. They aim to show that separate models should be trained depending on the activity of the user on the phone, as well as the movement of the user. Similar to our research, [19] used a dataset from online, the HMOG dataset. From the touchscreen data they extracted the mean, variation, percentiles, time, length, velocity, direction, gesture shape related features, as well as the start and end point of the swipe in x and y direction. From the touch pressure signal the mean, variation, percentiles, and shape related features are computed. Features obtained from the accelerometer data are calculated 0.5 seconds before and after a swipe. The features extracted from this data were the mean, minimum, maximum, variation, and percentiles. As an attempt to identify the features that are the best to choose for authentication, the model finds the mean of all features from the genuine user, and the mean of all features from the imposter, and finds the features with the greatest distance from each other. The highest-ranking features are then chosen to be used for training in a distance based one-class classifier. [19] tested their models in varying combinations of reading or navigating and walking or sitting. They found that while the user was sitting, touch screen data performed very slightly better on its own, but when the user began walking, combining touchscreen and phone movement data drastically improved results. When the user was walking in one scenario, the touchscreen produced an EER of 21.7%, but when the phone movement data was added that dropped to an EER of 13.2% [19]. This study further reiterates that using a multi-modal system is beneficial for authentication models.

Researchers in [20] develop a multi-modal model that requires users to enter in a random ten-digit code in order to obtain behavioral and touchscreen data from the user. Similarly to [19], this work examines the effect walking, sitting, going down stairs, and standing has on their model. The model captures data using the gyroscope and accelerometer. Both of these sensors are three dimensional, and the model adds a fourth dimension by calculating the magnitude. The mean, standard deviation, skewness, and kurtosis are calculated from each of the dimensions, providing a total of 16 features from each sensor. 38 time based features extracted from the touchscreen data are also used in this model. The features are then fused together leaving a 70 feature vector representing each user. Then using a Weka-based feature selection scheme - the Information Gain Attribute Evaluator (IGAE) - the features are ranked accordingly. A threshold is calculated by dividing the number of features by the number of users in the dataset, and that ratio determines how many features will be kept. [20] found that the motion-based features dominated the touch-based features. Bayesian Networks (BN), K-Nearest Neighbors (KNN), Multilayer Perceptron (MLP) and Random Forest (RF) were applied in this research. After applying 5-fold cross validation, the True Acceptance Rate (TAR) and the False Acceptance Rate (FAR) were calculated. TAR shows the rate at which users were correctly accepted, and FAR calculates the rate at which users are wrongly accepted. The models all had their best TAR while the user was sitting, with TARs of 69.37%, 70.05%, 71.84%, and 72.11% for BN, RF, MLP, and KNN respectively. Interestingly, the models had their worst TAR's while the user was walking, but their best FAR. By applying parameter optimization and IGAE, the results dramatically increased by nearly 15% in some cases, with the TAR being 85.77% in sitting scenarios [20]. This study provides further evidence that multi-modal authentication schemes are promising.

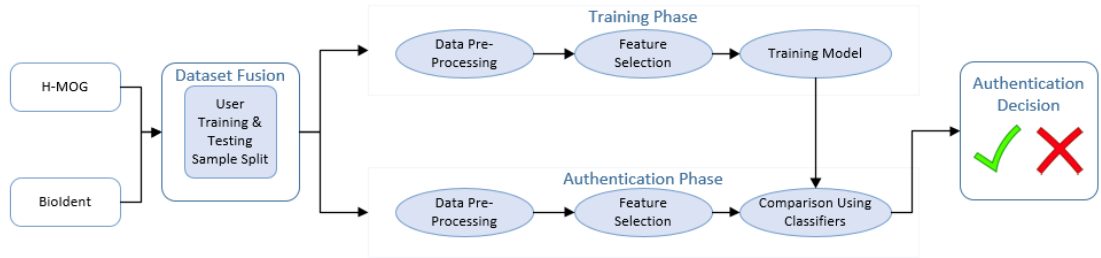

Figure 1.  Framework for our proposed model

III. METHODOLOGIES

A. System Overview

Our framework as seen above in Fig. 1, consists of four main phases: Dataset Fusion, Training Phase, Authentication Phase, and Authentication Decision. The first phase, Dataset Fusion, takes in the data from two publicly available datasets: BioIdent and H-MOG. In this phase the datasets are fused together into one, and then for each user there is a training and testing dataset created for them. Using an 80/20 train-test split for each user we take 80 genuine samples and 80 impostor samples for training and then for testing we use the remaining 20 genuine samples and all of the impostor samples. After these datasets are created for each individual user, the training datasets go to the training phase and the testing datasets go to the authentication phase. In both the training and authentication phases the data is pre-processed and then feature selection takes place. As we discuss below, our dataset consists of 24 chosen features. In the training phase the model is then trained after the feature selection to create an authentication signature for each user. In the authentication phase the authentication signature created in the training phase and the new data from the training dataset are used along with our three classifiers discussed below to compare the new data with the authentication signature of the user. This outputs the authentication decision of being either a genuine user or an impostor.

B. Dataset Description

For our dataset, we combined the data from two publicly available datasets: the H-MOG dataset and the BioIdent dataset. The H-MOG dataset was collected by yeng et al. [21], and the BioIdent dataset was collected by Antal et al. [22]. The reason for combining these two datasets was to get the phone movement features from the H-MOG dataset and add them to the touch dynamic features of the BioIdent dataset. The H-MOG dataset includes data from 100 different users over 24 sessions, but for our model we tested using one session from 51 different users. The BioIdent dataset includes data from 100 users collected in multiple sessions, and we used 100 samples from 51 users from that set. All together we ended up with 25 different raw features: user_id, stroke_duration, start_x, start_y, stop_x, stop_y, direct_end_to_end_distance, mean_resultant_length, up_down_left_right, direction_of_end_to_end_line, largest_deviation_from_end_to_end, average_direction, length_of_trajectory, average_velocity, mid_stroke_pressure, mid_stroke_area_covered, acc_x, acc_y, acc_z, gyro_x, gyro_y, gyro_z, mag_x, mag_y, mag_z. The first 15 features come from the BioIdent dataset and the last nine come from the H-MOG dataset. The user_id feature is our target for our classifiers. The stroke_duration is the total time taken to complete the stroke movement. The startX, startY, stopX, and stopY, correspond with the x-coordinate and y-coordinate at the start of the movement and the x and y coordinates at the end of the movement. The direct_end_to_end_distance is the direct length of the stroke between the stop and start points. The mean_resultant_length measures the amount of curve in the stroke. The up_down_right_left feature uses displacement to find the orientation of the stroke. The average_direction is the average slope for the stroke trajectory. The length_of_trajectory is the overall length of the stroke. The average_velocity is the stroke's average velocity. The mid_stroke_pressure is the pressure at the stroke's midpoint. The mid_stroke_area is the area that is covered by the finger at the stroke's midpoint [22]. The acc_x, acc_y, and acc_z corresponds with the x, y, and z coordinates of the accelerometer respectively. The gyro_x, gyro_y, and gyro_z corresponds with the x, y, and z coordinates of the gyroscope. Finally, the mag_x, mag_y, and mag_z are the x, y, and z coordinates of the magnetometer [21].

TABLE I. USER-SPECIFIC RESULTS FOR USER 45

| Success Metrics | Classifiers | | |
|---|---|---|---|
| | *RF* | *SVM* | *KNN* |
| Accuracy | 84.29% | 82.86% | 72.86% |
| Precision | 64.52% | 62.5% | 51.28% |
| Recall | 100% | 100% | 100% |
| F1 Score | 78.43% | 76.92% | 67.8% |
| EER | 11% | 12% | 19% |

Overall, our dataset ended up being just over 5,100 data samples. We used a training testing split of 80/20 for each individual users' data. Therefore, each user-specific training set consisted of 80 genuine user samples and an equal 80 impostor user samples which were taken equally from the other 50 users. Some users had more than one sample taken from them, but not enough to create a significant imbalance. For testing, the 20 remaining samples from the genuine user were used, along with a single sample from each of the remaining 50 users to act as impostor samples.

### C. Classifiers

To test the performance of our model we used three different classifiers. These classifiers were Random Forest (RF), Support Vector Machine (SVM), and K-Nearest Neighbor (KNN). RF is an ensemble classifier that uses multiple different decision trees to "vote" on a classification decision. SVM is a discriminative classifier that creates an n-dimensional space of multiple hyperplanes. Then from these multiple hyperplanes, one is determined to be the most optimal, and then its distance from new data points is used in its classification decision. Finally, KNN is a lazy classifier similar to SVM. However, in this algorithm, there is a user-specified number of "neighbors" whose distances from a new sample are used to "vote" on the classification of the new data.

## IV. EVALUATION RESULTS

### A. Success Metrics

To measure the performance of our model we used the following evaluation metrics: accuracy, precision, recall, F1 Score, and Equal Error Rate (EER). Accuracy is simply just a measure of the percentage of correctly classified user samples. Precision measures how precise our model is, as it finds the percentage of true positives out of all of the predicted positives our model created. Recall is a similar measurement to precision as this uses true positive cases to find the sensitivity of our model. Recall measures the percentage of true positives out of all samples that were supposed to be deemed positive, thus recall can also be considered our True Acceptance Rate (TAR). F1-Score is similar to accuracy however, it uses precision and recall to calculate its percentage. This means that any true negative samples are not being added into the percentage, so it is only looking at the effect the false negatives and false positives have on our model's effectiveness. Finally, EER is the point at which our False Acceptance Rate (FAR) and our False Rejection Rate (FRR) are equal. We want this value to be low, because the lower the EER the higher the effectiveness of our model.

### B. Results

Overall, our model showed promising results. Since we trained and tested each user individually, to look at overall effectiveness we calculated averages for each of the five success metrics. An example of the results found for a single user can be seen in Table 1. Overall, Random Forest was the best performing algorithm with an average accuracy of 81.74%. The other two algorithms also performed moderately overall, however; Random Forest consistently produced significantly higher accuracies when looking at individual users. These averages can be seen in Table 2. As seen in Table 2, the recall of our model for each algorithm was consistently high, with the lowest average being 89.4% for SVM and the highest being with RF with 97.35%. We also produced relatively low EERs for each algorithm, ranging from 13.56% to 20.68%. We also saw modest results for both F1-Score and precision, both of which can be seen in Table 2. We also produced the ROC curve for each algorithm for each user, an example of a ROC curve for single user can be seen in Fig. 2. Overall, the ROC curves produced were in the acceptable and excellent test quality range, with extremely few area under the curve (AUC) scores being below 0.7 and multiple being over 0.9.

## V. DISCUSSION AND ANALYSIS

As mentioned earlier, RF was the best performing algorithm with SVM and KNN coming in second and third respectively. This is consistent with the findings in [17]. These results may also be due to the fact that the dataset we were working with was relatively small. As stated in [17], SVM often times is less effective with smaller datasets, which could explain its weaker performance in this study in comparison to other studies such as [13, 23-25].

As mentioned previously, our dataset had just over 5,100 data points, which totaled out to 100 data points per user. While the training dataset for each user consisting of 160 samples was balanced, the testing set for each user was also very small, including only the twenty remaining samples from the genuine user and a single sample from each user to act as 50 impostor samples. This difference in amounts of genuine and impostor samples did create a slight imbalance in our testing dataset, which could slightly skew the data. However, it is important to note that slight imbalances are very common in real world scenarios, so the slight imbalance seen in our model is nothing of great concern and our results can be interpreted just as they would normally as the slight imbalance did not cause a significant change in outcome of our results.

TABLE II. AVERAGE RESULTS FOR ALL USERS

| Success Metrics | Classifiers | | |
|---|---|---|---|
| | *RF* | *SVM* | *KNN* |
| Accuracy | 81.74% | 76.98% | 73.63% |
| Precision | 63.06% | 57.75% | 53.21% |
| Recall | 97.35% | 89.4% | 92.55% |
| F1 Score | 75.34% | 68.19% | 66.23% |
| EER | 13.56% | 19.21% | 20.68% |

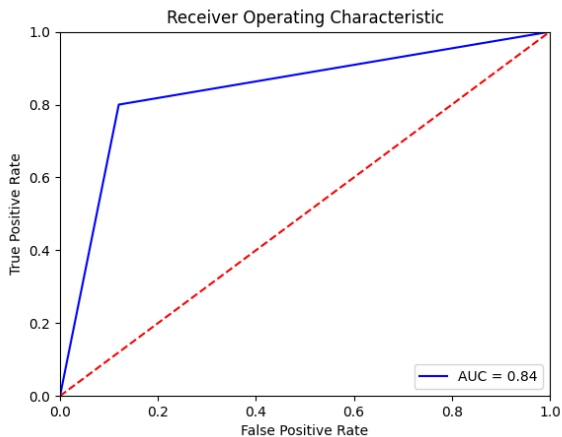

Figure 2. Example of ROC Curve taken from user 8's SVM Performance

Another significant finding from these results were the high recall averages along with the moderate precision averages. The high recall scores are promising, as it shows that our model is accurately classifying the genuine users and very rarely does it deny a genuine user access by classifying it as an imposter. This is incredibly important for the usability of this model. Since the model will rarely deny user's access, users are going to feel confident in using this schema and will not be inconvenienced by a faulty authentication schema. However, our model produced only moderate precision averages. Having only a moderate precision indicates that our model is inaccurately classifying impostor users as genuine relatively frequently. While the averages were all still above 50%, which suggests that our model still does a relatively good job at denying impostor users, when it comes to a security standpoint, having false acceptances is the exact opposite of what we would like from our model. Looking at both the precision and recall together, it would suggest that our model simply just not as selective as we would like. Again, this could be due to the smaller dataset, as having a larger dataset would allow our model to have more examples of what a genuine user should look like in training. However, it is also important to note that it is hard to get both a high precision and a high recall, as it is difficult to find a perfect balance between the two. Working to increase our precision could reduce our recall, which would reduce our usability. However, in a real-world scenario, most users would likely prefer higher security over higher usability.

## VI. LIMITATIONS

While our authentication scheme showed promise, there are several different variables that could be affecting our results. Firstly, our dataset was limited in size. Even after fusing two datasets together, this was still an issue for our model. This issue could cause consistency problems with our model and affect our accuracy. Having a model that is inconsistent can result in accuracies appearing better or worse than they actually are. Another issue we had that arose from fusing two datasets together was the inability to ensure that the new users being created from the two datasets were being fused with data from two similar users. If one user is highly knowledgeable in their use of technology and the other user is someone with very little experience in technology, that could create inaccurate data for our newly created users. Finally, our accuracies could be affected by the datasets not specifying between single and multi-touch datapoints. This can affect accuracies positively, making the results seem better than truly are.

## VII. CONCLUSION AND FUTURE WORK

This paper introduced a multi-modal authentication model utilizing multi-modal behavioral biometrics, specifically touch dynamics and phone movement. We used 2 publicly available datasets, H-MOG and BioIdent, that were captured utilizing the touchscreen, accelerometer, gyroscope, and magnetometer sensors. After preprocessing the data and performing feature extraction, our model applied multiple machine learning algorithms, namely Random Forest, Support Vector Machine, and k-nearest neighbors. We were able to achieve an accuracy of 81.74%, which came from the Random Forest classifier.

Overall, our results display the promise behavioral biometrics has in authenticating a user on mobile devices. However, there is still work needed to be done to make this schema adequate for real world scenarios. Again, we must look at the balance between usability and security, we need to find the perfect balance between those two needs to ensure that our schema will be applicable in real world scenarios. We will continue to improve this model by looking to create a new, more robust, dataset than the publicly available ones currently, emphasizing the need to distinguish single and multi-touch gestures as this seems to plague many other research endeavors along with our own. With this new dataset distinguishing between single and multi-touch our model could have improved accuracies and more consistent representation.

## ACKNOWLEDGMENT

Funding for this project has been provided by the University of Wisconsin-Eau Claire's Karlgaard Computer Science Scholarship Foundation and the University of Wisconsin-Eau Claire's Office of Research and Special Programs Student-Faculty Collaboration Grant.